\documentclass{aa}

\DeclareRobustCommand{\TUSSEN}[3]{#2}
\usepackage{graphicx}
\usepackage{txfonts}
\usepackage{color}
\usepackage{hyperref}
\usepackage{natbib}

\begin{document}

\title{Lossless compression of simulated radio interferometric visibilities}

\author{A. R. Offringa\inst{1,2} \& R. J. van Weeren\inst{3}}

\institute{ASTRON, the Netherlands Institute for Radio Astronomy, Oude Hoogeveensedijk 4,7991 PD Dwingeloo, The Netherlands\\
    \email{offringa@astron.nl}
  \and
    Kapteyn Astronomical Institute, University of Groningen, PO Box 800, 9700 AV Groningen, The Netherlands
  \and
    Leiden Observatory, Leiden University, PO Box 9513, 2300 RA Leiden, The Netherlands}
\date{Date accepted: 26 December 2026}

  \abstract
   {Processing radio interferometric data often requires storing forward-predicted model data. In direction-dependent calibration, these data may have a volume an order of magnitude larger than the original data. Existing lossy compression techniques work well for observed, noisy data, but cause issues in calibration when applied to forward-predicted model data.}
   {To reduce the volume of forward-predicted model data, we present a lossless compression method called Simulated Signal Compression (Sisco) for noiseless data that integrates seamlessly with existing workflows. We show that Sisco can be combined with baseline-dependent averaging for further size reduction.}
   {Sisco decomposes complex floating-point visibility values and uses polynomial extrapolation in time and frequency to predict values, groups bytes for efficient encoding, and compresses residuals using the \textsc{deflate} algorithm. We evaluate Sisco on diverse LOFAR, MeerKAT, and MWA datasets with various extrapolation functions. Implemented as an open-source Casacore storage manager, it can directly be used by any observatory that makes use of this format.}
   {We find that a combination of linear and quadratic prediction yields optimal compression, reducing noiseless forward-predicted model data to 24\% of its original volume on average. Compression varies by dataset, ranging from 13\% for smooth data to 38\% for less predictable data. For pure noise data, compression achieves just a size of 84\% due to the unpredictability of such data. With the current implementation, the achieved compression throughput is with 534 MB/s mostly dominated by I/O on our testing platform, but occupies the processor during compression or decompression. Finally, we discuss the extension to a lossy algorithm.}
   {}

   \keywords{Instrumentation: interferometers; Methods: observational; Techniques: interferometric; Surveys; Radio continuum: general}

   \maketitle
   \nolinenumbers
%

\section{Introduction}

Modern radio astronomical observatories produce vast amounts of visibility data, for which the complex values are typically stored as two 32-bit single-precision floating-point values. During the processing of such data, it is often necessary to store model data alongside the observed visibilities: these are forward-predicted values that simulate the response of a telescope given a sky model and are commonly used in two processing steps: i) (self-)calibration, where gains are solved to minimize the difference between the model visibilities and observed data; and ii) deconvolution, where they are used to compute the residual image in Cotton-Schwab-based techniques \citep{cotton-schwab-clean}. Such data are of the same size as the observed visibility data.

Storing and transporting large data volumes is a major expense for observatories, which will be even greater for the Square Kilometre Array (SKA) currently under construction. The large size of the model data can be particularly problematic for multi-stage calibration approaches \citep{degasperin-2019} or for direction-dependent self-calibration \citep{vweeren-2016}. In the latter case, it may be necessary to store a model column for each calibrated direction \citep{dejong-2024}, which can reach about ten times the size of the observed data.

The observed visibilities are noisy and can be compressed using lossy compression techniques. Dysco is a compression technique that automatically finds appropriate scaling for interferometric data and performs non-linear quantization and bit-packing \citep{offringa-dysco-2016}. With the Dysco compression technique, interferometric data can be compressed by approximately a factor of four with an insignificant increase in noise \citep{offringa-dysco-2016,chege-dysco-2024}. By using non-linear quantization optimized for Gaussian-distributed data, it produces visibilities for which the sum of squared residuals is minimal for a given compression factor.

The lossy compression method offered by the FITS format is aimed at imaging data \citep{pence-2010-fits-fp-compression} and is based on linear quantization and Rice encoding. More generic approaches also exist, such as the MultiGrid Adaptive Reduction of Data (\texttt{mgard}) technique \citep{gong-mgard-2023}, which employs detrending, linear quantization, and Huffman encoding. This method has also been applied in the context of radio interferometric data \citep{dodgson-et-al-2025}, and was shown to provide good compression with insignificant impact on the image quality. The \texttt{mgard} technique does not offer an automated way to perform scaling appropriately for radio interferometric data, in which different correlated visibilities may be formed from antennas with varying noise levels—a particularly significant effect in very long baseline interferometry (VLBI) — and currently requires manually setting an absolute or fractional error bound.

Although the aforementioned compression techniques for noisy visibility data are available, applying quantization-based compression techniques with similar compression rates to model data adds noise to the otherwise noise-free model data. This causes problems in processing, particularly in calibration, where the observed data is calibrated using these model data. This makes lossy compression unsuitable for model data, and in this paper we therefore explore lossless compression.

Model data are noiseless and generally smooth in the time and frequency directions. Although off-axis sources can reduce the smoothness of the data, the integration time and frequency resolutions are typically chosen to minimize smearing on the longest baselines, ensuring that the visibility data remain smooth on all shorter baselines. This effect is further mitigated by the attenuation of these distant sources by the primary beam. Many compression algorithms can exploit correlations between adjacent values, and a common approach for compressing such data is to predict the next value based on previous value(s) and compress the difference using an entropy encoder such as Rice-Golomb encoding \citep{golomb-1966,rice-encoding-1971} or Huffman encoding \citep{huffman-1952}. One well-known example of such an algorithm is the Free Lossless Audio Codec (FLAC; \citealt{flac-2024}), which compresses the difference between adjacent values using Rice encoding. However, FLAC is designed for compressing integer values, whereas visibilities are floating-point values, making the encoding not directly applicable. The FITS format supports lossless compression using Rice compression for image data but also supports only integer data in lossless mode \citep{pence-2009-lossless}.

Generic lossless algorithms, such as GZIP (which employs the \textsc{deflate} mechanism combining LZ77 and Huffman coding), are widely used due to their balance of speed and compression ratio. However, they do not exploit the spectrally and temporally smooth behaviour of radio astronomical model data that can be leveraged for better performance.

Another method to reduce the volume of interferometric data is baseline-dependent averaging (BDA; \citealt{wijnholds-bda-2018}). This can reduce data by a factor of a few but comes in most cases at the cost of some information loss: reducing the time and frequency resolution of the data may complicate or prevent data calibration or experiments requiring high time or frequency resolution, such as the search for radio signatures of star-planet interactions or exoplanets (e.g., \citealt{callingham-2024}), fast-radio bursts (e.g., \citealt{pleunis-lofar-frb-2021}), gamma-ray bursts (e.g., \citealt{rowlinson-2024}), spectral lines (e.g., \citealt{heywood-mightee-2024}) and polarized emission at non-zero Faraday depth (e.g., \citealt{osullivan-2023}).

In this paper, we develop a novel lossless compression framework for radio astronomy data that, for the first time, targets  noiseless model data. It integrates prediction for floating-point data and compresses this with the \textsc{deflate} algorithm. By leveraging predictive preprocessing to reduce entropy, followed by the efficient dictionary-based and Huffman encoding of the off-the-shelf \texttt{libdeflate} library\footnote{\href{https://github.com/ebiggers/libdeflate}{https://github.com/ebiggers/libdeflate}}, our approach achieves significantly improved compression ratios over generic data compression algorithms. This algorithm is especially suitable for interferometric model data, as the memory footprint per baseline is minimal, thereby allowing on-the-fly compression. The algorithm is released with an open-source license and integrated into Casacore\footnote{\href{https://github.com/casacore/casacore}{https://github.com/casacore/casacore}}, allowing all existing software that uses Casacore Measurement Sets to use the compression technique, providing transparent compression for users. The DP3 and WSClean software packages include an option to write data using Sisco compression\footnote{Sisco compression can be enabled with the ``\textsc{-model-storage-manager sisco}'' option in WSClean and with ``\textsc{msout.storagemanager=sisco}'' in DP3.}, and support both regular and BDA data. When software uses the Casacore library, any Sisco-compressed data are transparently decompressed during access. Apart from enabling the compression, no further manual settings are required.

We describe the details of the method in Section 2 of this paper. In Section 3, we present results of applying the method in various scenarios and on various test sets. Finally, Section 4 discusses the results and conclusions.

\section{Method}
Sisco consists of four stages: i) prediction of a data value based on previous values; ii) subtracting the predicted value by decomposing both the floating-point data value and the predicted value into mantissa, exponent, and sign values, shifting the mantissa of the predicted value to the same exponent, and subtracting their mantissas; iii) grouping the resulting data so that all exponent bytes and each of the four mantissa with sign bytes are close together; and iv) applying the \textsc{deflate} algorithm separately to the exponent values and the combined sign and residual mantissa values.

We describe each of these steps in more detail in the following sections.

\subsection{Predicting a data value}
For each data value, we perform two-dimensional polynomial extrapolation of the real and imaginary values in both the time and frequency dimensions. Implementation-wise, this is computed by performing two one-dimensional extrapolations. We implement and test prediction using 0th to 3rd order polynomials, as well as compression without any prediction.

Zeroth-order extrapolation results in simply compressing the difference with respect to the previous value, which is commonly used (e.g., in FLAC; \citealt{flac-2024}). First, second, and third-order extrapolation imply linear extrapolation using the last two values, quadratic extrapolation using the last three values, and cubic extrapolation using the last four values, respectively. This results in the following prediction values:
\begin{eqnarray} \label{eq:order0}
 p^0_i &=& x_{i-1} \\
 p^1_i &=& 2x_{i-1} - x_{i-2} \\
 p^2_i &=& 3x_{i-1} - 3x_{i-2} + x_{i-3} \\
 \label{eq:order3}
 p^3_i &=& 4x_{i-1} - 6x_{i-2} + 4x_{i-3} - x_{i-4}
\end{eqnarray}
where $p^a_i$ is the predicted value of order $a$ for data value $i$, and $x_{i-b}$ is the $b$-th previous value before sample $i$.

To analyze whether adding more values to the extrapolated fit would improve compression, we also test using the average of two values, linear prediction from three values, and quadratic prediction from four values:
\begin{eqnarray}
 q^0_i &=& \frac{1}{2} \left( x_{i-1} + x_{i-2} \right) \\
 q^1_i &=& \frac{1}{3} \left( -2 x_{i-1} + x_{i-2} + 4x_{i-3} \right) \\
 \label{eq:last-order}
 q^2_i &=& \frac{1}{4} \left( 9x_{i-1} - 3x_{i-2} - 5x_{i-3} + 3x_{i-4} \right)
\end{eqnarray}
respectively. The prediction equations all follow from performing a least-squares fit on the values for the given polynomial and extrapolating these values to the next value in the sequence. These equations are applied first in the time direction, after which the equations are applied again on the residuals, but now in the frequency direction. For the results, we generally apply the same equation in the time and frequency directions. However, to assess whether a combination of equations could be effective, we also include results in which we apply linear prediction in time and quadratic prediction in frequency. We also test compression without prediction.

We selected these prediction methods instead of a multi-grid approach that relies on hierarchical multi-resolution decomposition, because our prediction requires only a few preceding samples in the time direction and no additional per-baseline state --- an important advantage given the specific data ordering of interferometric data  (see \S\ref{sec:reordering}).

The visibility weights are not used in the extrapolation. While these could improve the prediction, it involves significantly more operations to perform a weighted extrapolation. Visibilities that are close together in time and frequency generally have similar weights, so a weighted fit would likely be similar to a non-weighted fit. Flagged visibilities that are effectively zero-weight visibilities might show more erratic patterns, particularly if they result from interference detection. Since flagged values are typically a small percentage of the total data, we decided that it was not worth doing so. Bad predictions due to, e.g., flagged RFI-containing visibilities will result in a larger variation of the residual values, which are then harder to compress. This decreases the compression rate, but the compression remains lossless.

In our implementation, we perform the equations \eqref{eq:order0}--\eqref{eq:last-order} using integer math by first shifting all the previous values such that they have the same exponent as $x_i$, i.e., the currently compressed data value. After that, the prediction calculations are performed on the mantissa, which is stored as an unsigned 32-bit integer. This gives precise control over how to round when shifting a mantissa to increase the exponent and allows capturing 32-bit mantissa values in intermediate calculations. While using double-precision values for the calculations would also provide such accuracy and avoid the implementation of integer-based floating-point manipulations, the decomposition and manipulation of the floating-point values are already required in the next steps, so in the full algorithm, this speeds up the calculations while providing full rounding control.

In the special case where not enough previous values are available to perform the prediction of the selected order, or if some of them are not-a-number (NaN) or $\pm$ infinity, a lower order is selected until sufficient values are available. If no values are available at all, the prediction is skipped, and the data value is directly compressed. The same approach is used when overflow occurs during the prediction.

\subsection{Forming the residual}
During compression, the residual data value is calculated by subtracting the predicted value from the data value, and during decompression, the predicted value is added back to the residual value. These operations are performed by matching the exponents of the predicted value to that of the data value and performing integer math on the mantissa. Unlike the previous step, where integer math is not crucial, here it is required to ensure that $c_i = x_i - p_i$ is exactly reversible using $x_i' = p_i + c_i$, which is not necessarily true for floating-point operations. For example, when $x_i$ is so much smaller than $p_i$ that shifting its exponent truncates the mantissa (or makes it zero), the reconstructed value $x_i'$ becomes truncated (or zero). This is avoided by always shifting the exponent of the predicted value toward the data value. This may cause truncation of the predicted value, but since this happens during both compression and decompression, it is not an issue.

Shifting the exponent of the predicted value during decompression requires knowing the exponent of the data value before decompressing it. This requires storing the (unshifted) exponent of the original data along with the sign and mantissa of the residual. Together, they form the residual value in the base of the original data. To compress the data effectively, the residual values should be small, hence requiring low encoding entropy. The exponents are also likely to cover only a small number of different values. The sign of the residual can be expected to be uniformly distributed and, therefore, incompressible.

In the exceptional case where the compressed value is the special value NaN or $\pm$ infinity, the predicted value is not subtracted, and the 'residual' value is set to the data value. During decompression, such a value can be determined from the separately stored exponent value, as NaN and infinity values use the reserved exponent value of +128.

Similarly, if the exponent value of the residual is larger than that of the data value, indicating a bad prediction, the value is also stored directly. During decompression, this can be identified by comparing the exponent of the predicted value with the stored exponent.

\subsection{Grouping the data} \label{sec:grouping}
After subtracting the residual as described in the previous section, the data to compress is a stream of signs, mantissas, and exponents of the residuals. We use a 32-bit integer to store the mantissa and the sign. The 32nd bit is used for the sign, and the other 31 bits for the mantissa. If the 32nd bit (or higher bits) of the mantissa would have been set by the calculations in the prediction, it is considered an overflow, causing the mantissa of the data value itself to be stored instead of the residual.

Before calling \textsc{deflate}, the data are concatenated into five separate groups: one group holds the exponent bytes, a second group holds the first byte of the mantissa with the sign bit, and the third to fifth groups hold the second, third, and fourth mantissa bytes. Finally, the five groups are concatenated into one chunk. This grouping ensures that similar values are close together in the data stream offered to \textsc{deflate}, and we found that this substantially improves compression.

The data are chunked to limit the memory required for large datasets. We empirically determined that a million samples is a reasonable compromise between memory usage and compression efficiency.

\subsection{Applying \textsc{deflate}}
\textsc{deflate}, described in \citet{deflate-rfc}, uses a combination of Lempel-Ziv 1 dictionary encoding \citep{lempel-ziv-1977,lzss-1982}, which is effective in compressing repeated sequences, and Huffman encoding \citep{huffman-1952}, which uses shorter encoding symbols for words that occur more often. The algorithm is widely used and is an integral part of \textsc{zip} compression. It is asymmetric in its speed: compression is significantly slower than decompression. The implementation we use, \textsc{libdeflate}, aims to make use of modern hardware features and provides a tunable compression level from 1 to 12, where 1 is the fastest but with the worst compression, and 12 is the slowest with the best compression.

Once a million samples have been collected, the \textsc{deflate} library is called to compress the grouped data chunk, which is subsequently written to disk along with a small header containing identification, versioning information, the prediction method, and compressed and uncompressed sizes.

\subsection{Data ordering considerations} \label{sec:reordering}
Measurement sets are typically ordered by timesteps (i.e., it is the slowest-changing dimension when serialized) and then by baseline. Most operations, including calibration and imaging, need to process the data in this order. For calibration, having the full bandwidth for a single timestep close together is important for applying smoothing or regularization over frequency \citep{mitchell-mwa-2008,yatawatta-2015,gan-2023}. For simple imaging, the order of the data is less critical, but having the data in time order can be practical for example when the beam is applied during imaging \citep{tasse-aprojection-2013,van-der-tol-idg-2018}. The detection of radio-frequency interference is an exception and often processes the data ordered by baseline \citep{offringa-apertif-rfi-2023}. Sisco uses the correlation in the time and frequency directions of visibilities of the same baseline, and to overcome the issue that the time direction is spaced apart, it is necessary to keep a buffer of the last few timesteps with all baselines, polarizations, and spectral channels in memory while the data are written or retrieved. Because the number of previous timesteps required is low (three timesteps for quadratic prediction), the memory required for this is insignificant with modern computing facilities. This buffering mechanism allows Sisco to be transparently applied on data with the standard measurement set order. The computational cost associated with the data buffering and reshuffling is a fraction of the total computational cost of compressing, and is included in the computational performance tests of \S\ref{sec:computational-performance}.

Since the compression and decompression of any visibility require a number of previous visibilities, which themselves require decompression of their previous visibilities, and so on, it follows that: i) to decompress a visibility, all its previous visibilities must be decompressed; and ii) to change a visibility, all subsequent visibilities must be recompressed. It would be possible to work around this issue by independently compressing chunks of a certain size, where each chunk assumes no previous state. However, most data access is done by streaming from beginning to end and not randomly through the file, so we have chosen simplicity and not implemented this. As a result, when the data are accessed randomly, our implementation becomes inefficient.

\subsection{Baseline-dependent averaging}
Baseline-dependent averaging (BDA) is a technique to reduce data size by averaging in time and frequency while using different time and frequency averaging factors for different baselines \citep{wijnholds-bda-2018}. Because many interferometers have relatively many short baselines, which can be averaged by large factors, this technique can substantially reduce the number of visibilities, although the averaging factors are often constrained to maintain sufficient temporal and spectral resolution for calibration gain corrections or for science cases, such as radio exoplanet searches \citep{callingham-2024} or to preserve spectral lines (e.g., \citealt{ciardi-21cm-forest-2015, oonk-2016-rrl}). In practice, for LOFAR observations, the total achieved reduction in size from using BDA is often on the order of a factor of two to three when combined in both the time and frequency directions.

Because baselines are processed independently with the Sisco technique, it is relatively easy to combine with BDA. Since Sisco makes no use of the time grid of a baseline, supporting time-direction BDA requires no particular changes, but frequency-direction BDA does. In a measurement set, frequency BDA is supported by reusing the spectral window table to associate each baseline with the correct list of channels for the averaging factor used. Each measurement set row may then hold a data array with a different size. In Sisco, the data are flattened and written sequentially to disk, and the shape of each array is stored in a separate file. This file is also compressed using the \textsc{deflate} algorithm.

\section{Results}
We apply Sisco to a variety of datasets to demonstrate its compression performance and determine settings that work well. We start with simulated noiseless data, for which Sisco is intended, but also apply Sisco to observed data to measure its performance on noisy data. For all results, we have verified that the compressed and decompressed data are bit-by-bit identical to the original data, i.e., the compression is truly lossless.

\begin{table*}[htb]%
\caption{Scenarios used for evaluating Sisco. The listed size is the size of the uncompressed data for a single direction.  }%
\label{tbl:scenarios}\begin{center}\vspace*{-5mm}\begin{tabular}{|l|c r r c c|}
\hline 
 & & & & & \textbf{Unique} \\
 & \textbf{Target} & \textbf{Frequencies} & \textbf{Size} & \textbf{Generated by} & \textbf{correlations} \\
\hline
MWA & Vela & 134--164 MHz & 0.8 GB & WSClean (Image based) & 4 \\
MeerKAT & Abell 2933 & 0.6--1.1 GHz & 123 GB & WSClean (Image based) & 1 \\
LOFAR HBA & Abell 1795 & 120--167 MHz & 8.4 GB & WSClean (Image based) & 1 \\
LOFAR LBA & 3C48 & 59 MHz & 14 GB & DP3 (Analytic) & 4 \\
\hline
\end{tabular}
\end{center}
\end{table*}

\subsection{Compressing a simple sine wave}
To investigate the behaviour of Sisco in a simple scenario, we compress data produced by the following function:
\begin{equation} \label{eq:sinewave}
 V(t, f) = \exp \left(\alpha it + \beta if\right).
\end{equation}
Here, $t$ and $f$ are the time and frequency indices, respectively, and $i$ is the imaginary unit. The constants $\alpha$ and $\beta$ are set to values with twelve random decimals, where $\alpha \approx 0.13$ and $\beta \approx 0.17$. This produces a simple two-dimensional sine wave in the time and frequency directions, with a reasonable (but arbitrary) period and without trivial repetitions of values. This function has the property that the higher-order prediction functions of Eqs.~\eqref{eq:order0}--\eqref{eq:last-order} will be more accurate in predicting the next value. While this may not be realistic for real data, it allows us to analyse the compression effectiveness when prediction is accurate. As input, we evaluate the function for a grid of 10,000 $\times$ 10,000 values, resulting in a data volume of $8 \times 10^8$ bytes, or $\sim 763$ MB.

\begin{figure}
  \begin{center}
	\includegraphics[width=9.3cm]{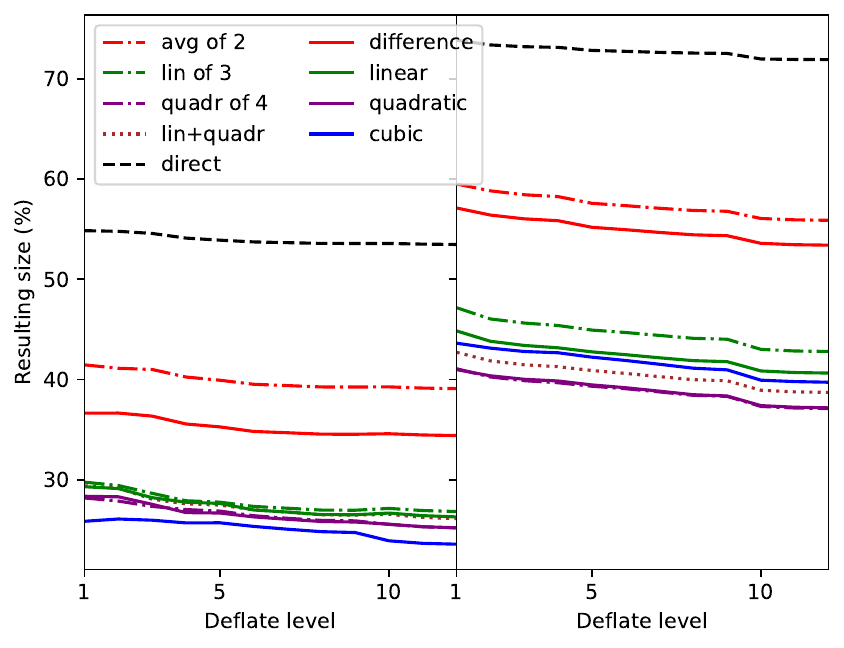}%
	\caption{Left panel: Compression results when using a two-dimensional sine wave as input. Right panel: Compression result for noiseless forward-predicted MWA data. Both plots show the compression results for various prediction schemes and as a function of the \textsc{deflate} level.}
	\label{fig:simulated-sin-compression}
  \end{center}
\end{figure}

As a first reference, we compress the file with the floating-point data stored as binary values using the standard \texttt{zip} and \texttt{bzip2} software packages with their best compression settings. The resulting file sizes are 91.4\% and 91.5\%, respectively, highlighting that, even though it is generated data, such standard compression algorithms are not very effective in reducing the data size. Fig.~\ref{fig:simulated-sin-compression} shows the Sisco compression results for the different prediction methods and as a function of the \textsc{deflate} level. By using Sisco without prediction, the data are compressed to a resulting size just below 55\%. The difference between \texttt{zip} and Sisco with \textsc{deflate} is remarkable since \textsc{deflate} is the algorithm used by \texttt{zip}. This difference results from grouping the data before \textsc{deflate} (described in \S\ref{sec:grouping}). Enabling prediction further improves compression considerably. The best compression is achieved with cubic prediction and the highest \textsc{deflate} level, resulting in a size of 23.6\%. In this configuration, 99.3\% of the compressed data is occupied by the mantissa and sign data, and 0.7\% by the exponent data, demonstrating that the exponent data is easily compressed.

\begin{figure*}
  \begin{center}
	\includegraphics[width=9cm]{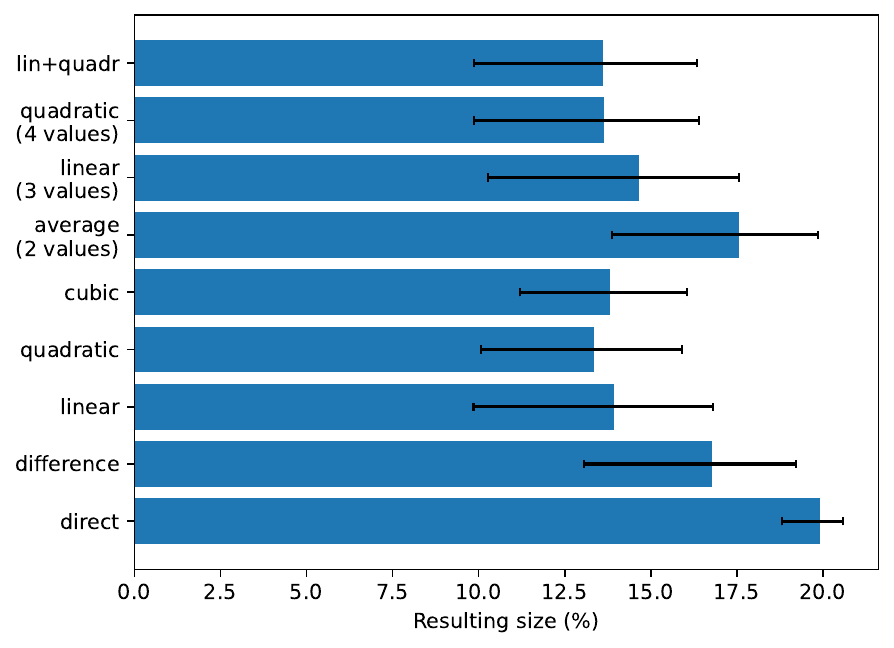}%
	\includegraphics[width=9cm]{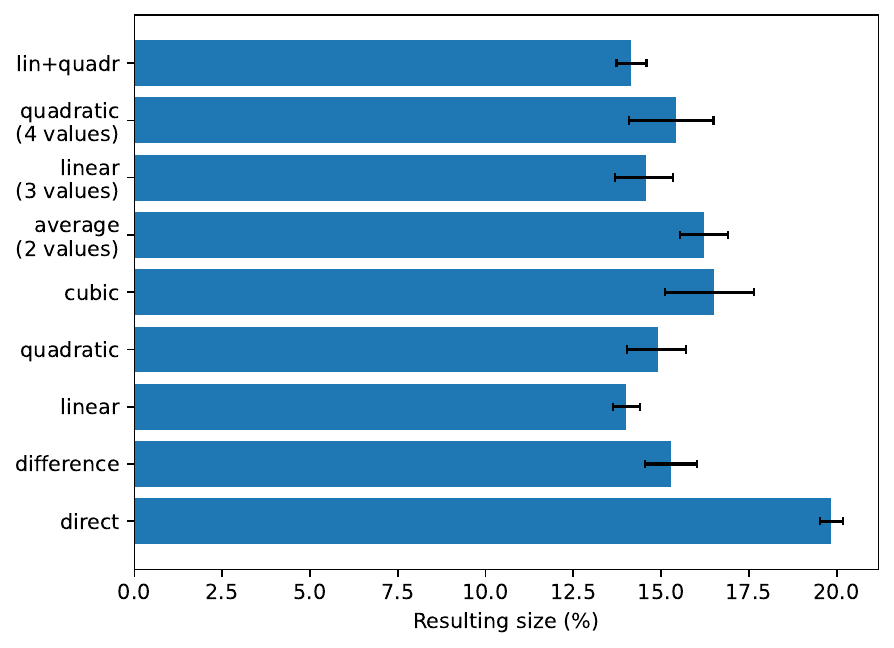}%
	\caption{Compressed size of simulated noiseless visibility data for MeerKAT (left plot) and LOFAR HBA (right plot) with various prediction methods. In both cases, forward-predicted models were compressed for a number of different directions. The MeerKAT set has 11 directions, and the LOFAR HBA set has 5 directions. Error bars are drawn from the minimum to the maximum compression results.}
	\label{fig:meerkat-lofar-model-compression}
  \end{center}
\end{figure*}

\subsection{Forward-predicted data compression results}
We apply Sisco to the scenarios listed in Table~\ref{tbl:scenarios}. In this section, we use forward-predicted data without noise. While these forward predictions use $uvw$-values and other metadata from a real observation, any flags due to RFI or missing data are ignored during the forward prediction, i.e., 100\% of the data is predicted and compressed.

For the MWA data set, the simulated data were produced by running WSClean \citep{offringa-2014-wsclean} in joined polarization mode to compute clean components and fill the model data with a polarized model from those. A relatively large image size of 50 degrees was imaged to include sources far from the phase center.  Compression results for this simulated MWA set are shown in the right panel of Fig.~\ref{fig:simulated-sin-compression} for different prediction modes and as a function of the \textsc{deflate} level. We find that, when using \textsc{deflate} level 9, the best compression of 38.4\% is achieved with quadratic prediction, compared to 72.5\% without prediction. The \textsc{deflate} level setting has relatively little effect on the compression: quadratic prediction with levels 1, 9, and 12 produces compression results of 40.1\%, 38.4\%, and 37.2\%, respectively. Although the difference in compression speed between levels 1 and 9 is insignificant, the use of \textsc{deflate} level 12 makes Sisco 2.5 times slower. With \textsc{deflate} level 9 and quadratic prediction, the 24 bits in the original data for the mantissa and sign are compressed to 64\%. As can be expected, the 8-bit exponent data are much more compressible, and compress to 15\% of their original size.

Fig.~\ref{fig:meerkat-lofar-model-compression} shows, in the left plot, the compression results for the MeerKAT dataset. This dataset contains data for 11 forward-predicted directions, and a \textsc{deflate} compression level of 9 is used. The forward-predicted data in this set contain the XX, XY, YX, and YY correlations. However, all XX and YY values are the same, and the XY and YX values are zero as a result of predicting Stokes I values into the dataset without full Jones (beam or calibration) corrections. Therefore, simply storing only a Stokes I value instead of all four correlations would shrink the data to 25\%. Of the methods tested, quadratic prediction achieves the best compression of 13.3\%. Adding more values to the fit by using the average of two, linear fit of three, or quadratic fit of four produces slightly worse results compared to their counterparts with the minimum number of values in the fit. Direct compression of the visibilities without prediction provides the worst compression of 19.9\%, achieving only a small additional compression.

For the MeerKAT data, we find again that the 8-bit exponent data can be compressed much more than the 24-bit combined mantissa and sign data. With \textsc{deflate} level 9 and quadratic prediction, the exponent data of the first direction are compressed to just 3\% of their original size, and the mantissa and sign data to 17\%, resulting in a total compressed size of 10.1\%. We use these same data to analyze whether it would be more effective to manually store and compress a single Stokes I value instead of letting the compression algorithm remove the redundancy. We find that compressing XX, XY, YX, and YY individually results in a compressed data volume of 2.64~GB, whereas a single Stokes I value results in 2.53~GB, an additional 5\% saving. Compressing only Stokes I values makes the compression 2.3 times faster.

The right panel of Fig.~\ref{fig:meerkat-lofar-model-compression} shows the results for the LOFAR HBA dataset. The five directions in this set yield average compression results broadly similar to those obtained for MeerKAT, except that in this case, linear prediction outperforms quadratic prediction. The LOFAR HBA dataset is representative of direction-dependent LOFAR use-cases where model visibilities are stored --- for instance, when deriving direction-dependent solutions for the Dutch LOFAR stations, which are essential for achieving high-quality results in the widefield VLBI use-case (see \citealt[Sect. 3.3.5]{dejong-2024}). Other examples include obtaining direction-dependent calibrations in widefield mode for both HBA and LBA observations with the Dutch array, as well as for smaller extracted regions \citep{vweeren-2021}. Similar use-cases occur for other instruments such as MeerKAT (e.g., \citealt{botteon-2024}) and the GMRT, where direction-dependent calibration is likewise essential. In all these cases, Sisco is highly beneficial due to the large size of intermediate data products and the large number of directions that must be predicted and stored prior to calibration.

\begin{figure}
  \begin{center}
	\includegraphics[width=9cm]{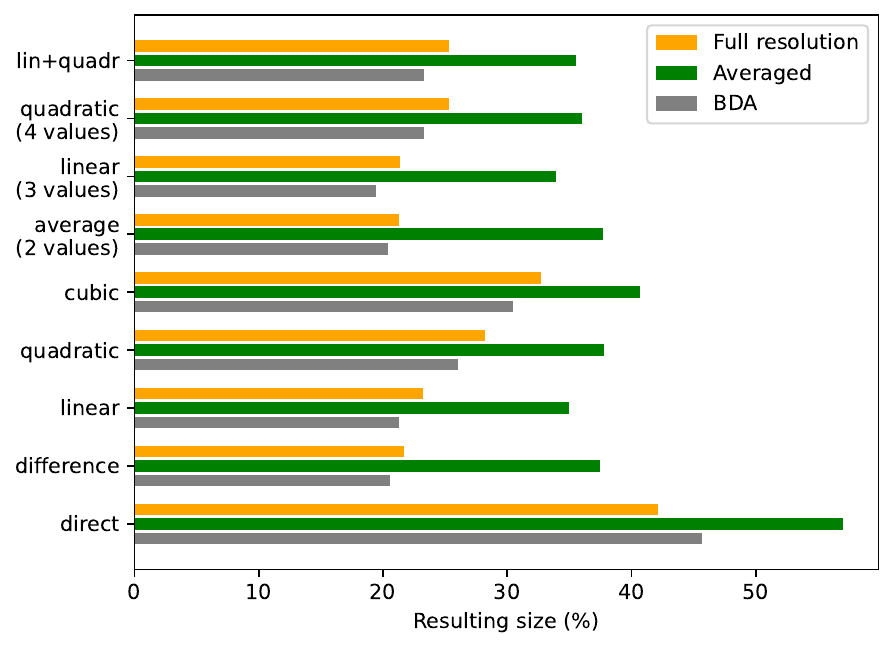}\\%
	\caption{Compression results for noiseless forward-predicted LOFAR LBA data with high time and frequency resolution. }
	\label{fig:lofar-3c48-model-compression}
  \end{center}
\end{figure}

Fig.~\ref{fig:lofar-3c48-model-compression} shows the compression results for a LOFAR LBA simulated dataset with very high time and frequency resolution and with only a source in the center, as is typical for a forward-predicted calibrator observation. This is an extreme example in which the values are extremely slowly varying in time and frequency. During simulation, the LOFAR beam was applied, and the XX, XY, YX, and YY correlations are therefore not redundant in this case due to the projection effects of the fixed LOFAR dipoles. In this set, compressing with prediction from the previous value, the average of two values, and linear prediction from three values achieves the best compression, all with similar compression results of 21.7\%, 21.3\%, and 21.4\%, respectively. Turning prediction off performs the worst, achieving a compression of 42.1\%. Averaging the data by a factor of 4 in frequency and 8 in time before compression results in a lower compression rate by Sisco, which highlights that quicker variations in the data make the data less compressible. For the averaged data, linear prediction from three values produces the best compression result of 33.9\%. Fig.~\ref{fig:lofar-3c48-model-compression} also shows the results of compressing the same dataset after BDA. For the averaging, we used the DP3 software with a time and frequency base of 500~km, ensuring that baselines longer than 500~km are not averaged. This averaging reduces the data by a factor of 2.6. The Sisco compression rates on BDA results are on average 2.7\% better than the full-resolution results. The reason for the similar behaviour between BDA and full-resolution data is that the long, unaveraged baselines dominate the average level of compression, because BDA will not lower their resolution. Overall, the results demonstrate that BDA and Sisco combine well, reaching a total reduction in size of over a factor of 13. With the BDA data, linear prediction from three values achieves the best compression result of 19.5\%.

\begin{table*}[htb]
\caption{Compression results for forward-predicted data. All values are the resulting fractional size in percentages. Each result uses \textsc{deflate} level 9. The MeerKAT and LOFAR HBA values are averaged over all directions. The LOFAR LBA values are the average over the full resolution, time and frequency averaged, and BDA results. The best compression value for each dataset is printed in bold.}
\label{tbl:all-results}
\begin{center}
\begin{tabular}{|l|ccccc|}
\hline 
\textbf{Prediction method} & MWA (\%) & MeerKAT (\%) & LOFAR HBA (\%) & LOFAR LBA (\%) & Average (\%) \\
\hline
No prediction & 
72.5 & 19.9 & 19.8 & 48.3 & 40.1 \\
Previous value & 
54.3 & 16.8 & 15.3 & 26.6 & 28.3 \\
Linear & 
41.8 & 13.9 & \textbf{14.0} & 26.5 & 24.1 \\
Quadratic &
38.4 & \textbf{13.3} & 14.9 & 30.7 & 24.4 \\
Cubic &
41.0 & 13.8 & 16.5 & 34.6 & 26.5 \\
Average from 2 values & 
56.8 & 17.6 & 16.2 & 26.5 & 29.3 \\
Linear from 3 values & 
44.0 & 14.7 & 14.6 & \textbf{24.9} & 24.6 \\
Quadratic from 4 values & 
\textbf{38.3} & 13.6 & 15.4 & 28.2 & \textbf{23.9} \\
Linear in time, quadratic in frequency & 
39.9 & 13.6 & 14.1 & 28.0 & \textbf{23.9} \\
\hline
\end{tabular}
\end{center}
\end{table*}

A summary of the results is given in Table~\ref{tbl:all-results}, including the averages over all datasets. The datasets show small differences in the effectiveness of the prediction schemes. Linear and quadratic prediction show the best compression levels. Averaged over all sets, the combination of linear prediction in the time direction and quadratic prediction in the frequency direction shows the best results. For sets other than the LOFAR LBA set, the difference between this combined prediction scheme and the best prediction method is within 1\%. Given that the LOFAR LBA is a somewhat extreme example due to its high resolution, the linear-quadratic combined prediction is likely a good all-around default setting for compression with Sisco.

\subsection{Observed data compression results}

\begin{table}[htb]%
\caption{Sisco compression results on observed (and therefore noisy) data. The listed values are the resulting fractional size after compression in percentages. \textsc{deflate} level 9 was used. The last column lists the percentage of flagged visibilities, which has a substantial effect on compression.}%
\label{tbl:noisy-data-results}%
\begin{center}%
\begin{tabular}{|lccc|}
\hline 
Test set & No predict (\%) & Linear (\%) & Flags (\%) \\
\hline 
MWA & 59.1 & 64.9 & 40 \\
MeerKAT & 53.1 & 57.7 & 62 \\
LOFAR HBA & 79.4 & 86.7 & 40 \\
LOFAR LBA & 84.1 & 92.3 & 0 \\
\hline
Gaussian data & 84.1 & 92.3 & 0 \\
\hline
\end{tabular}%
\end{center}%
\end{table}

While not the primary aim of this compression technique, in this section, we investigate the compression efficiency of Sisco for observed (and therefore noisy) data.

Table~\ref{tbl:noisy-data-results} lists the results without prediction and with linear prediction for the actual observed data from all test sets of Table~\ref{tbl:scenarios}, as well as for pure Gaussian noise. We find that the use of prediction consistently worsens compression. This can be expected from noise-dominated data, as extrapolation from a low number of visibilities will increase the noise and make the predicted value further from the observed value. 

The first three test sets have a significant number of flags due to failing stations and—for the LOFAR data—flagging of bright sources in the sidelobes that occurs in pre-calibration processing as described in \citet{morabito-vlbi-lofar-2022}. Due to the repeated null values in the data, higher flag ratios make the data more compressible. Predictionless compression of the LOFAR LBA set, which contains no flags, results in 84.1\%, which is identical to the result for simulated Gaussian data. The standard deviation of the data is approximately 250~Jy, whereas the target source 3C48 has an average apparent flux density of approximately 100~Jy in this observation. Averaging the LOFAR LBA data does not change the compression ratios. Applying the \texttt{zip} and \texttt{bzip2} algorithms with the best compression settings on the Gaussian data results in sizes of 92.5\% and 95.0\%, respectively. The increased compression achieved by Sisco comes again from grouping the data as described in \S\ref{sec:grouping}.

\subsection{Computational performance} \label{sec:computational-performance}

We use a commodity-level AMD Ryzen 7 2700X Eight-Core Processor from 2018 for benchmarking our implementation of Sisco. On generated data, Sisco achieves a compression throughput of 534~MB/s at \textsc{deflate} level 9. The use of real data from disk lowers this to 340~MB/s due to the extra I/O involved in reading the data. The computational cost of Sisco is mostly dominated by the \textsc{deflate} algorithm, though the different prediction levels do have a small impact on the compression speed. The more data values are used, the slower the prediction. The use of \textsc{deflate} levels 10-12 decrease the throughput considerably, with a throughput of 61~MB/s and 50~Mb/s at level 12 with generated and real data, respectively. Given that the extra compression achieved at level 12 is not substantial compared to level 9, we conclude that level 9 is a good default level for the use in Sisco.

Solid-state drives (SSDs) can achieve write speeds of several GB/s, which means that Sisco may slow down writing when using fast SSDs. On the other hand, the use of Sisco will increase the write speed on slower drives because it lowers the amount of data written. For LOFAR, the raw visibility stream before any averaging is done can be a few GB/s \citep{lofar-2013}, which, if this were to be compressed with Sisco, would require 5-10 nodes that would compress the data concurrently. After interference detection, the LOFAR data stream is averaged down by as much as a factor 16, after which the data stream could be compressed in real-time on a single node.

Since the prediction of baselines is independent of each other, and the \textsc{deflate} compression of chunks is also independent, our implementation performs the independent steps concurrently on multiple cores. Many use-cases combine data reading and writing with computationally intensive tasks, such as calibration or gridding. While this can lower the wall-clock time of I/O related overhead, the computational power required for Sisco may compete with the other processing tasks, and Sisco may thus lower the wall-clock time of such processing tasks. Nevertheless, the computational requirements for Sisco are relatively minor compared to calibration and imaging tasks.

\section{Conclusions and discussion}
We have presented a lossless compression algorithm for forward-predicted radio interferometric model data. By using a combination of data grouping, prediction and the application of \textsc{deflate}, we achieve an average compression factor higher than 4. We found that prediction using a linear, quadratic or combined function provides the best compression results, giving an average resulting size of approximately 24\%. This is comparable to the compression rate achieved when Dysco would be applied to forward-predicted data, with the benefit that it is lossless. With a compression and decompression speed faster than traditional spinning disk drives, the extra cost involved for Sisco is likely not an issue in processing pipelines. Moreover, by its integration into the Casacore library, it can be seamlessly integrated into existing pipelines.

\citet{dodgson-et-al-2025} use \texttt{mgard} for visibility compression, which uses a multi-grid approach. To reduce buffer sizes, we use prediction using polynomial extrapolation. In general, polynomial extrapolation performs well when the higher-order derivatives of the signal are small (i.e., the data are smooth), whereas multi-grid or multi-resolution methods excel at capturing large-scale trends even in the presence of sharp local features.

In Sisco, we have made use of the \textsc{deflate} library. This allows us to reuse an already optimized entropy and dictionary encoding library. Entropy encoding is crucial to compress the residuals with frequent small values. Dictionary encoding, however, is not often used on scientific data, and since we have not investigated the effectiviness of Huffman and Lempel-Ziv (LZ) encoding separately, it may be unnecessary and needlessly increase the computational intensity of Sisco. \citet{fauzan-lzw-2022} conclude that, compared to the Lempel-Ziv-Welch algorithm, Huffman encoding is much more computationally demanding, and in a trade off between computing cost and compression achieved, it may be worthwhile.

While compression and decompression increases the wall-clock time of processing tasks, we find the benefit of reduced storage requirement already outweigh the computational costs in for example the LOFAR widefield VLBI science case. So far, we have only made a simple CPU implementation. In similar compression algorithms, the use of GPUs has been shown to provide a substantial increase in speed \citep{knorr-2021, lossless-gpu-compression-2025}. The data order size of interferometric data sets may form challenges in doing so, but solutions may become available as more operations, such as calibration and imaging, are moved to GPUs.

\subsection{Sisco as a lossy method}
It is relatively simple to extend Sisco to perform lossy instead of lossless compression and let it work on noisy data. This can be achieved by clearing a number of least significant bits of the residual mantissa in combination with adding jitter to avoid systematic errors. This would result in an algorithm with a relative error bound per visibility, similar to \textsc{mgard} \citep{gong-mgard-2023}. For radio astronomical visibility data, such a bound is however not particularly useful, because visibilities are normally not individually used. Rather, they are averaged over baselines, time and frequency, for example when forming an image. With a relative error bound, small values will have a smaller absolute error, even though these have a smaller effect on the average. Moreover, a number of quantization levels are reserved for particularly large values, that are very unlikely to occur in Gaussian distribution data. Lossless and lossy algorithms require therefore by nature different approaches to achieve good results.

Dysco takes the expected distribution into account and performs an encoding that, for a given compression level, results in a least-squares error on the average \citep{offringa-dysco-2016}. However, Dysco does not exploit correlations or repetitions in the data,  which Sisco can. Moreover, Dysco can also not be applied on BDA data. There is therefore room to combine beneficial properties of the two to construct a lossy algorithm that works on BDA data, achieves optimal least-squares residuals on the average over Gaussian data and can also exploit correlations in the data.


\bibliographystyle{aa}
\DeclareRobustCommand{\TUSSEN}[3]{#3}
\bibliography{sisco-bibliography}

\end{document}